\documentstyle[psfig]{l-aa}

\begin{document}

   \thesaurus{           
              (13.25.5 
               08.14.1;  
               08.16.7)} 
   \title{Spin-down rate of 1E 2259+586 from
RXTE observation }

   \author{A. Baykal
          \inst{1}, J.H. Swank \inst{2},
 T. Strohmayer \inst{2}, M.J. Stark \inst{2}
          }

   \offprints{Altan Baykal }

   \institute{ 
  Physics Department, Middle East Technical University,
  Ankara 06531, Turkey \inst{1}, 
Laboratory for High Energy Astrophysics
NASA/GSFC
Greenbelt, Maryland 20771 USA \inst{2}}

   \date{Received March 23, 1998; accepted April 17 , 1998}

   \maketitle
  \markboth{Baykal et al.,}{Baykal et al.,}

   \begin{abstract}

We present new X-ray observations of the X-ray pulsar 1E 2259+586, obtained
during March 1997, with the Rossi X-Ray Timing Explorer (RXTE). We have
measured the pulse frequency derivative $\dot \nu = (-1.08 \pm 0.04) \times
10^{-14}$ Hz s$^{-1}$ from pulse arrival times 
obtained in a sequence of 5 observations
spread over one month. This $\dot\nu$ is consistent with the long term
spin-down trend. We also found that the observed X-ray luminosity is
consistent with that measured at quiescent X-ray flux levels by previous
missions. Our observations imply that 1E 2259+586 was spinning down
steadily without exhibiting any stochastic torque noise fluctuations
during the month covered by our observations.

\end{abstract} 
\section{Introduction}

The X-ray pulsar 1E 2259+586 is located at the geometric center of the
semi-circular shell of diffuse X-ray emission of the supernova remnant
G109$-$1.0 (Gregory \& Fahlman 1980).  Radio observations indicate that the
age of G109$-$1.0 is about $10^4$ yr and the distance to it is 3.6--4.7 kpc
(Hughes et al. 1984). The pulse frequency of the pulsar is 0.1433 Hz (pulse
period 6.8 sec) and the average spin down rate since its discovery is $\dot
\nu = -1 \times 10 ^{-14}$ Hz s$^{-1}$.  If 1E 2259+586 is an isolated,
spinning-down pulsar, this rate yields a rotational energy loss of
10$^{31}$ erg s$^{-1}$ which is much less then the intrinsic X-ray
luminosity of $\sim $10$^{35}$ erg s$^{-1}$ (Corbet 1995).  Koyama et
al., (1987, 1989) claimed that the spin-down rate and unabsorbed X-ray
luminosity can be understood if 1E 2259+586 is in a binary system with a
neutron star magnetic field of 5$\times 10 ^{11}$ Gauss and spinning close
to the equilibrium period (Ghosh \& Lamb 1979).  However, efforts to find
orbitally induced Doppler shifts in the X-ray pulsations which would reveal
the binary period have been unsuccessful.  The upper limits of $a_{x} /c
\sin i$ from previous missions such as Einstein, EXOSAT, Ginga and ROSAT
have been very small.  For example, recent RXTE observations (Mereghetti et
al., 1998) have given the lowest upper limit $a_{x}/c \sin i < 30$ msec at
the 99\% confidence level.  According to these new upper limits, if 1E
2259+586 is in a binary system, its companion star must be either a white
dwarf, or a helium-burning star with $M < 0.8 M_{\circ}$ (Mereghetti et
al. 1998). Alternatively, 1E 2259+586 could be an isolated star which is
accreting from a disk and is formed
by remnants of the common envelope evolution
of a high-mass X-ray binary (van Paradijs et al. 1995, Ghosh et al. 1997).

Analyses of the ROSAT observations (Baykal \& Swank 1996) have shown that
the source is not steadily spinning-down.  A spin-up episode which is
superposed on a long term spin-down trend was found. The deviations from
the secular spin-down trend are also consistent with other accretion
powered X-ray pulsars (Baykal \& {\"O}gelman 1993), which are several orders
of magnitude greater than those of radio pulsars. This result favors
accretion onto a neutron star from a very low mass companion or from a
residual disk. In this research note, we present the pulse timing results
and X-ray spectrum of 1E 2259+586 from RXTE observations.  The main goal of
these new RXTE observations is to resolve the pulse frequency derivative
$(\dot \nu)$ during the observation time span and measure the source
X-ray luminosity. These quantities enable us to further test the current
accretion hypothesis for this source.

\section{Observation and Data Analysis}

1E 2259+586 was observed between 1997 February 25 and March 26 on 5
different days separated from each other by approximately one week.  Each
25 ksec observation was taken in approximately one day.  The results
presented here are based on data collected with the Proportional Counter
Array (PCA, Jahoda et al. 1996). The PCA instrument consists of an array of
5 proportional counters operating in the 2--60 keV energy range, with a
total effective area of approximately 7000 cm$^2$ and a field of view,
$\sim 1^{\circ}$ FWHM.

\subsection{Torque and X-ray Luminosity}
 
Background light curves and the pulse height amplitudes are generated using
background estimator models based upon the rate of very large events (VLE),
spacecraft activation and cosmic X-ray emission with the standard PCA
analysis tools and are subtracted from the source light curve obtained from
the Good Xenon event data. After correcting the background subtracted light
curves with respect to the barycenter of the solar system, data sets in
each observation were folded on statistically independent trial periods
(Leahy et al. 1983). A master pulse was constructed by folding the data on
the period giving the maximum $\chi^2$. The master pulse with 55 phase
bins was represented by its Fourier harmonics (Deeter \& Boynton 1985) and
cross-correlated with the harmonic representation of average pulse profiles
from each observation. The pulse arrival times so obtained are represented
in Fig. 1. The quadratic trend of pulse arrival times is a direct
measure of the change of the pulse frequency during the observation (or
intrinsic pulse frequency derivative),
\begin{equation}
\delta \phi = \phi_{o} + \delta \nu (t-t_{o})
+ \frac{1}{2} \dot \nu (t-t_{o})^{2}
\end{equation}
where $\delta \phi $ is the pulse phase offset deduced from the pulse
timing analysis, $t_{o}$ is the mid-time of the observation, $\phi_{o}$ is
the phase offset at t$_{o}$, $\delta \nu$ is the deviation from the mean
pulse frequency (or additive correction to the pulse frequency), and $\dot
\nu $ is the pulse frequency derivative of the source.  Table 1 presents
the timing solution of 1E 2259+586 from our RXTE observations.  The pulse
frequency derivative obtained from the quadratic trend of the pulse timing
analysis is $\dot \nu _{RXTE} = -(1.08 \pm 0.04)\times 10^{-14} $ Hz
s$^{-1}$ which is consistent with the average spin-down rate over a time
span of 19 years, $<\dot \nu > = -(1.15 \pm 0.06) \times 10^{-14}$ Hz
s$^{-1}$.  Fig. 2 presents the previous pulse frequency history of 1E
2259+586 together with the new RXTE observation.

\begin{table}
\caption{Timing Solution of 1E 2259+586 for RXTE Observations
   }
\label{Pri}
\[
\begin{tabular}{|c|c|}  \hline
Epoch(MJD)  & 50516.494628      \\ \hline
Pulse Frequency (Hz) &0.1432886474 $\pm $ 1.8$\times$10$^{-9}$  \\

Pulse Freq. Derivative (Hz s$^{-1}$)  &
$-(1.08 \pm 0.04) \times 10^{-14}$ \\ \hline
\end{tabular}
\]
\end{table}

\begin{figure}
\psfig{file=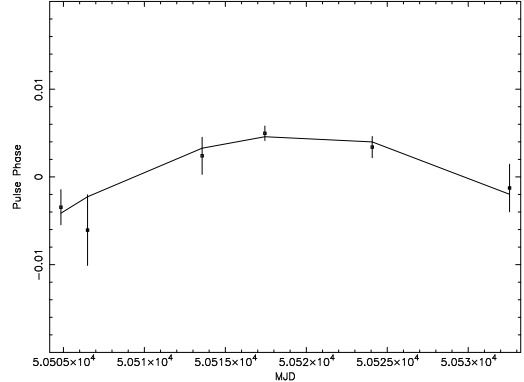,height=2.3in,angle=-90}
\caption[]{Phase ofsets in pulse arrival times.}
\end{figure}
\begin{figure}
\psfig{file=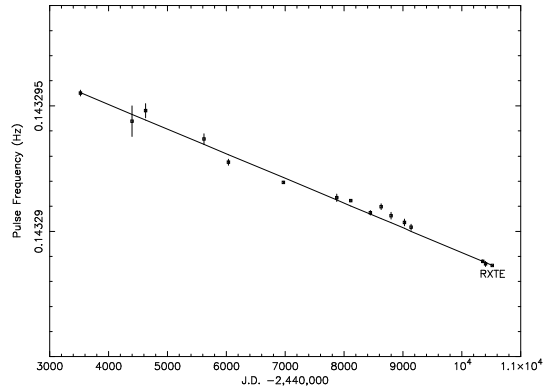,height=2.3in,angle=-90}
\caption[]{Pulse frequency history of 1E2259+586,
               including the latest RXTE
               measurement.}
\end{figure}
\begin{figure}
\psfig{file=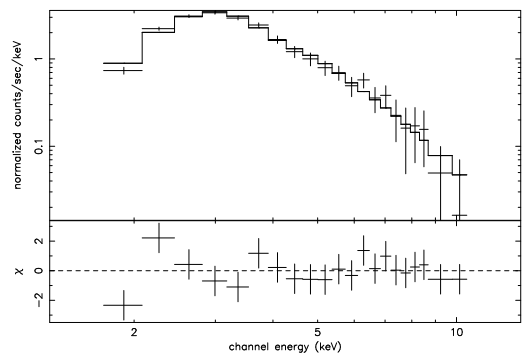,height=2.3in,angle=-90}
\caption[]{(top) RXTE PCA X-ray spectrum of 1E2259+586 with the
                 fit to the
                 power law model.
                 (bottom) The residuals to the fit. The reduced
                 $\chi ^{2}$ = 1.1.}
\end{figure}
\begin{figure}
\psfig{file=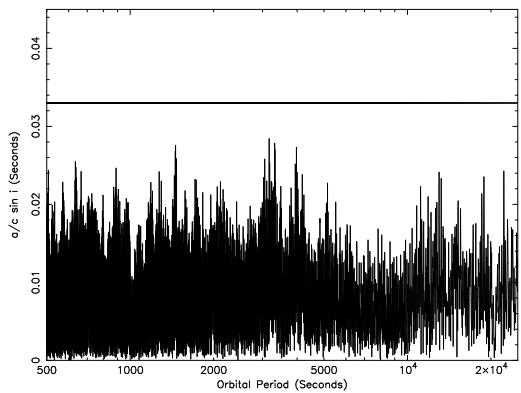,height=2.3in,angle=-90}
\caption[]{Results of a search for orbital Doppler
               delays for 1E 2259+586. The horizontal
               line denotes the
               a$_{x}$/c sin i values corresponding to detection at
               the 99$\%$ confidence level.}
\end{figure}
The background spectrum was calculated using the same background estimator
models based upon the rate of very large events (VLE), spacecraft
activation and cosmic X-ray emission as used to calculate background
light curves.
The resultant background subtracted spectrum in the energy range 2-10 keV  
 yields X-ray luminosity $L_{x}=(0.38 \pm
0.06)\times 10^{35} (\frac{d}{4{\rm kpc}})^{2}$ erg s$^{-1}$ which is
consistent with that obtained from ASCA and SAX measurements in 2-10 keV
(Corbet et al., 1995, Parmar et al., 1997).
 The best
fitting power law X-ray spectrum with column density N$_{H}$=(2.2$\pm$ 0.8)
$\times$ 10$^{22}$ cm$^{-2}$ and photon index 4.78 is presented in Fig. 3.
The background contamination from the supernova remnant is
only a few percent of the total source flux (Rho \& Petre 1997). ASCA and
Beppo SAX observations have found a soft black-body component with a
temperature of 0.44 keV (Corbet et al. 1995, Parmar et al. 1997).  Our fits
to the X-ray spectra do not resolve the soft black-body component since the
RXTE/PCA detectors are insensitive to energies less than about 2 keV.
There is no evidence for any deviation from a power law that might be
attributable to a cyclotron feature.

\subsection{Search for Orbital Period }

In searching for orbital Doppler modulation, orbital periods from $\sim
500$ sec to $\sim 2.5 \times 10^{4}$ sec were considered by oversampling
the independent Fourier periods (Leahy 1983) by a factor of 4.  For each
trial orbital period, 8 pulse profiles are obtained at the different
intervals of the orbital phase by correcting the light curve for the secular
spin-down rate and folding the light curve at the the pulse period.  The
resulting pulse profiles were cross-correlated with the master pulse, which
was obtained from all observations. To search for any sinusoidal modulation
in the 8 pulse arrival times, Fourier amplitudes of the arrival times were
computed and are presented as a function of the trial orbital period in
Fig. 4.  In the absence of an orbital signal, the expected distribution
of Fourier amplitudes is the $\chi^{2}$ distribution with 2 degrees of
freedom (van der Klis 1989). As seen in Fig. 4, no significant signal
detection is found at the 99\% confidence level.  The upper limit for any
undetected signal is estimated as $a_{x} \sin i < 0.028$ light-s, which is
very close to the earlier upper limit given by Mereghetti et al. (1998).
(We note that a slight improvement of the upper limit arises from the
better statistics due to the longer RXTE observation with respect to
previous observations) The above upper limits on Doppler shifts confirms
that if the 1E 2259+586 is in a binary system, then the companion star must
be either a white dwarf, or a helium-burning star (Mereghetti et al.,
1998).

\subsection{Torque and X-ray Luminosity Changes in  
Accretion Powered Disk Fed Binaries and 1E 2259+586}

In disk-fed systems, the accretion onto the neutron star is believed to be
from a Keplerian disk (Ghosh \& Lamb 1979). The torque on the neutron star
is given by
\begin{equation}
I\dot \Omega = n(w_{s})  \dot M~l_{K},
\end{equation}
where $l_{K} = (GMr_{o})^{1/2}$ is the specific angular momentum added by a
Keplerian disk to the neutron star at the inner disk edge $r_{o} \approx
0.5 r_{A}$, where $r_{A} \ = (2GM)^{-1/7} \mu ^{4/7} \dot M^{-2/7}$ is the
Alfven radius, $\mu$ is the neutron star magnetic moment, $n(w_{s}) \approx
1.4 (1-w_{s}/w_{c})/(1-w_{s})$ is a dimensionless function that measures
the variation of the accretion torque as estimated by the fastness
parameter $w_{s}=\Omega /\Omega _{K}(r_{o}) = 2 \pi P^{-1} G^{-1/2}
M^{-5/7} \mu ^{6/7} \dot M^{-3/7}$.  Here $w_{c}$ is the critical fastness
parameter at which the accretion torque is expected to vanish ($w_{c} \sim
0.35-0.85$ depending on the electrodynamics of the disk, Lamb 1989). In
this model, the torque will cause a spin-up if the neutron star is rotating
slowly ($w_{s}~<~w_{c}$) in the same sense as the circulation in the disk,
or spin-down, if it is rotating in the opposite sense (see Lamb 1991).
Even if the neutron star is rotating in the same sense as the disk flow,
the torque will spin-down the neutron star if it is rotating too rapidly
($w_{s}~>>~w_{c}$).  In such a model one should see positive correlation
between angular acceleration ($\dot \Omega $) and mass accretion rate
($\dot M$) if the disk is rotating in the same sense as the neutron
star. If the flow is from Roche Lobe overflow then the accreting material
carries positive specific angular momentum $l$, therefore it is hard to
imagine accretion flow reversals and hence the spin-up/down torques should
be correlated with mass accretion rate $\dot M$.

In Ginga observations the source had flux levels a factor of two higher
than average (Iwasawa et al. 1992).  This implied that the mass accretion
rate on the source is indeed variable and fluctuations in the spin-down
rate and possible spin-up trends should be expected (Baykal \& Swank
1996).  However, our RXTE observations found the source with an X-ray flux
consistent with the secular spin-down.  Therefore, it is quite natural to
find the secular spin-down rate from pulse arrival times, although this
observation gives the spin-down rate of the source over the shortest
observing interval since the discovery of the source.

Recent observations of accreting neutron stars have shown stochastic
spin-up/down trends on time scales from days to a few years (Bildsten et
al. 1997).  In intervals between stochastic changes disk-fed sources show
secular spin-up or down trends with lower values of noise strength (Baykal
1997). Some of the sources switch from spin-up to spin-down states without
showing great changes in their mass accretion rates (Bildsten et
al. 1997). These unusual behaviors led Baykal (1997) and Nelson et al.
(1997) to the possibility of retrograde circulation of accretion disks. GX
4+1 shows correlation between the X-ray flux and the spin-down rate
(Chakrabarty et al. 1997) which may suggest a retrograde accretion disk.
In the application of the above observational results and current accretion
theory to 1E 2259+586, the spin-down rate should decrease or it should
switch to spin-up at higher accretion rates (higher X-ray flux).  On the
other hand if the source has a very unusual counter-rotating accretion
disk, in the high state the spin-down rate should increase with respect to
the low state. Future X-ray observations at a high state could address this
important question.\\

\begin{acknowledgements}
A.B. thanks A.Alpar, H.{\"O}gelman, {\"U}. K{\i}z{\i}lo\u{g}lu,
for stimulating discussions and USRA for supporting a visit to the
GSFC.
\end{acknowledgements}

\end{document}